\documentclass[preprint,preprintnumbers,amsmath,amssymb]{revtex4}
\usepackage{amssymb,graphicx}
\usepackage{amsmath,bm}

\newcommand{\bea}{\begin{eqnarray}}
\newcommand{\eea}{\end{eqnarray}}

\begin{document}
\title{Painlev{\'e} singularity structure analysis of three component
Gross-Pitaevskii type equations}
\author{T. Kanna\footnote{ Electronic mail: kanna{\_}phy@bhc.edu.in.}}
\affiliation{Department of Physics, Bishop Heber College,
Tiruchirapalli-620 017, India   }
\author{K. Sakkaravarthi}
\affiliation{Department of Physics, Bishop Heber College,
Tiruchirapalli-620 017, India   }
\author{C. Senthil Kumar \footnote{ Electronic mail: senthil{\_}bdu@rediffmail.com.}}
\affiliation{Department of Physics, VMKV Engineering College,
Periaseeragapadi, Salem-636 308, India}
\author{M. Lakshmanan \footnote{  Electronic mail: lakshman@cnld.bdu.ac.in.}}
\affiliation{Centre for Nonlinear Dynamics, Bharathidasan
University, Tiruchirapalli-620 024, India}
\author{M. Wadati \footnote{Electronic mail: wadati@rs.kagu.tus.ac.jp.}}
\affiliation{Department of Physics, Faculty of Science,
 Tokyo University of Science, 1-3 Kagurazaka,
 Shinjuku-ku, Tokyo 162-8601, Japan}

\date{\today}

\begin{abstract}
In this paper, we have studied the integrability nature of a
system of three coupled Gross-Pitaevskii type nonlinear evolution
equations arising in the context of spinor Bose-Einstein
condensates by applying the Painlev\'e singularity structure
analysis. We show that only for two sets of parametric choices,
corresponding to the known integrable cases, the system passes the
Painlev\'e test.
\end{abstract}

\maketitle
\section{Introduction}
Integrable multicomponent nonlinear Schr\"odinger type equations
have attracted considerable current interest in soliton research.
Much focus has been paid to identify new integrable multicomponent
type equations due to their many faceted applications in different
fields of science such as nonlinear optics, Bose-Einstein
condensates, biophysics, plasma physics,
etc.$^{\footnotesize{1-5}}$. Painlev\'e singularity structure
analysis is one of the powerful tools to isolate and identify
integrable dynamical systems$^{\footnotesize{6-10}}$. This
procedure nicely complements other integrability tools like
inverse scattering transform (IST), infinite number of involutive
integrals of motion,  symmetries, B\"acklund transformations,
Hirota's bilinearization method, etc., to study the integrability
properties of nonlinear systems$^{\footnotesize{1,2}}$. By
applying the Painlev\'e test for integrability a class of
integrable coupled nonlinear Schr\"odinger (CNLS) type equations,
which arise in different physical contexts, has been identified$^{\footnotesize{11-16}}$.\\

In this connection, the system of CNLS equations in the presence
of confining potential becomes the coupled Gross-Pitaevskii (GP)
equations, governing the dynamics of two component Bose-Einstein
condensates$^{\footnotesize{17-19}}$. This kind of multicomponent
condensates can also be created with the mixture of two different
atomic species or by considering the hyperfine spin of atoms in
the presence of optical dipole traps$^{\footnotesize{20-22}}$. The
latter entities are the so-called
spinor Bose-Einstein condensates (BECs).\\

Spinor Bose-Einstein condensates of ultra cold atoms can be
created by liberating the hyperfine states by means of optical
trapping. Two component condensates have been realized in
$^{87}Rb$~(see Ref.~23) and also optically trapped three component
condensates were studied in Refs.~24-27. The evolution of the
spinor condensate wave functions is governed by the following set
of three-coupled nonlinear Schr\"odinger type
equations$^{\footnotesize{28}}$,
\begin{subequations}
\bea
i\hbar{\psi_{+1,T}}&=&-\frac{\hbar^2}{2m}{\psi_{+1,XX}}+(c_0+c_2)(|{\psi_{+1}}|^2+|{\psi_0}|^2){\psi_{+1}}\nonumber\\
&&+(c_0-c_2)|{\psi_{-1}}|^2{\psi_{+1}}+c_2{\psi^*_{-1}}{\psi^2_0},\\
i\hbar{\psi_{0,T}}&=&-\frac{\hbar^2}{2m}{\psi_{0,XX}}+(c_0+c_2)(|{\psi_{+1}}|^2+|{\psi_{-1}}|^2){\psi_0} \nonumber\\
&&+c_0{|\psi_0}|^2{\psi_0}+2c_2{\psi^*_0}{\psi_{+1}}{\psi_{-1}},\\
i\hbar{\psi_{-1,T}}&=&-\frac{\hbar^2}{2m}{\psi_{-1,XX}}+(c_0+c_2)(|{\psi_{-1}}|^2+|{\psi_0}|^2){\psi_{-1}}\nonumber\\
&&+(c_0-c_2)|{\psi_{+1}}|^2{\psi_{-1}}+c_2{\psi^*_{+1}}{\psi^2_0},
\eea \label{eqn1}
\end{subequations}
where $\psi_{\pm1,0}$'s are the wave functions of the three spin
components, $T$ is the time and $X$ denotes the spatial
co-ordinate. The effective one-dimensional coupling constants
$c_0$ and $c_2$ representing the mean field and spin exchange
interactions, respectively, are given by $c_0=
\frac{g_0+2g_2}{3}$, $c_2 = \frac{g_2-g_0}{3}$, where
$g_f=\frac{4\hbar
^2a_f}{ma_\bot^2}\left(\frac{1}{1-C\frac{a_f}{a_{\bot}}}\right),~
f=0,2$. Here $a_f$'s are the s-wave scattering lengths in the
total hyperfine spin channel $f$, $a_{\bot}$ is the size of the
transverse ground state, $m$ is the atomic mass and the constant
$C=-\zeta(1/2)\simeq 1.46$, where $\zeta$ is the Reimann
zeta-function. With the redefinition of $T=\hbar t$,
$X=\frac{\hbar}{\sqrt{2m}} x$ and transforming
$(\psi_1,\psi_0,\psi_{-1})\rightarrow(\psi_1,\sqrt{2}\psi_0,\psi_{-1})$,
we can rewrite Eq.~(\ref{eqn1}) in the standard form as
\begin{subequations}
\bea
i{\psi_{+1,t}} &=& -{\psi_{+1,xx}}+(c_0+c_2)(|{\psi_{+1}}|^2+2|{\psi_0}|^2){\psi_{+1}} \nonumber\\
&&+(c_0-c_2)|{\psi_{-1}}|^2{\psi_{+1}}+2c_2{\psi^*_{-1}}{\psi^2_0},\\
i{\psi_{0,t}}&=&-{\psi_{0,xx}}+(c_0+c_2)(|{\psi_{+1}}|^2+|{\psi_{-1}}|^2){\psi_0} \nonumber\\
&&+2c_0{|\psi_0}|^2{\psi_0}+2c_2{\psi^*_0}{\psi_{+1}}{\psi_{-1}},\\i{\psi_{-1,t}} &=& -{\psi_{-1,xx}}+(c_0+c_2)(|{\psi_{-1}}|^2+2|{\psi_0}|^2){\psi_{-1}} \nonumber\\
&&+(c_0-c_2)|{\psi_{+1}}|^2{\psi_{-1}}+2c_2{\psi^*_{+1}}{\psi^2_0}.
\eea \label{eqn2}
\end{subequations}
\indent We refer to Eq.~(2) as the three-component GP type
equations. The above system of equations has been solved by the
IST method and multicomponent bright and dark solitons have been
reported for specific choices of $c_0$ and
$c_2$$^{\footnotesize{29-31}}$. Now it is of interest to isolate
all the possible integrable models arising from Eq.~(\ref{eqn2})
for arbitrary choices of $c_0$ and $c_2$, which can be tuned
suitably through Feshbach resonance. For this purpose, we perform
a Painlev\'e singularity structure analysis to the above fairly
generalized system. It is also expected that besides BECs the
analysis will have wider ramifications in nonlinear optics.

This paper is arranged in the following manner. In section II, the
three steps involved in the Painlev\'e singularity structure
analysis, namely the leading order analysis of the Laurent
expansion in the neighbourhood of a non-characteristic singular
manifold, determination of the resonances (that is, the powers at
which arbitrary functions can occur in the Laurent expansion) and
analysis of the Laurent expansion for sufficient number of
arbitrary functions are carried out. It is shown that only for the
two specific parametric choices, namely (i) $c_2=0$ and (ii)
$c_0=c_2$ the system (2) passes the Painlev\'e integrability test.
The results are analyzed in the final section.

\section{Painlev\'e singularity structure analysis}
In order to perform the Painlev\'e singularity structure analysis
of Eq.~(\ref{eqn2}) the dependent variables $\psi_{\pm1}$,
$\psi_0$ and their complex conjugates are denoted as \bea
\psi_{+1} = a, \quad \psi_{+1}^* = b, \quad \psi_{-1} = m , \quad
\psi_{-1}^* = n , \quad \psi_{0} = p, \quad \psi_{0}^* = q.
\label{pan1}\eea
 Then Eqs.~(2) become
\begin{subequations}
\begin{eqnarray}
ia_t &=& -a_{xx} + (c_0 + c_2)\,(ab+2pq)a + (c_0 - c_2)\,mna + 2c_2n p^2,\\
-ib_t &=& -b_{xx} + (c_0 + c_2)\,(ab+2pq)b + (c_0 - c_2)\,mnb + 2 c_2m q^2,\\
im_t &=& -m_{xx} + (c_0 + c_2)\,(mn+2pq)m + (c_0 - c_2)\,abm + 2c_2b p^2,\\
-in_t &=& -n_{xx} + (c_0 + c_2)\,(mn+2pq)n + (c_0 - c_2)\,abn + 2c_2a q^2,\\
ip_t &=& -p_{xx} + 2c_0p^2 q + (c_0 + c_2)\,(ab+mn)p + 2c_2qam,\\
-iq_t &=& -q_{xx} + 2c_0q^2 p + (c_0 + c_2)\,(ab+mn)q + 2c_2pbn.
\end{eqnarray}
\label{pan2}
\end{subequations}
The Painlev\'e singularity structure analysis (of an analytic
polynomial differential equation) is carried out by seeking a
generalized Laurent expansion$^{\footnotesize{32}}$ for the
dependent variables
\begin{subequations}
\bea
&&a = \phi^{\alpha} \sum_{j= 0} a_j(x,t) \phi^j, \quad~~ a_0\neq 0,\\
&&b = \phi^{\beta} \sum_{j= 0} b_j(x,t) \phi^j, \quad\quad b_0\neq 0,\\
&&m = \phi^{\gamma} \sum_{j= 0} m_j(x,t) \phi^j,  \quad m_0\neq 0,\\
&&n = \phi^{\delta} \sum_{j= 0} n_j(x,t) \phi^j, \quad~~n_0\neq 0,\\
&&p = \phi^{\epsilon} \sum_{j= 0} p_j(x,t) \phi^j, \quad\quad p_0\neq 0,\\
&&q = \phi^{\omega} \sum_{j= 0} q_j(x,t) \phi^j,  \quad\quad
q_0\neq 0, \eea \label{lser}
\end{subequations}
\noindent in the neighbourhood of the non-characteristic singular
manifold $\phi(x,t) = 0$, with nonvanishing derivatives
$\phi_x(x,t)\neq 0$ and $\phi_t(x,t) \neq 0$.

\subsection{Leading order analysis}
The leading order behaviour of the solution is analyzed by
assuming the forms \bea a \approx a_0\phi^{\alpha}, \quad  b
\approx b_0\phi^{\beta}, \quad   m \approx m_0\phi^{\gamma}, \quad
n \approx n_0\phi^{\delta}, \quad   p \approx p_0\phi^{\epsilon},
\quad  q \approx q_0\phi^{\omega} \eea for the dependent
variables, where $\alpha$, $\beta$, $\gamma$, $\delta$, $\epsilon$
and $\omega$ are integers to be determined. After substituting
these forms into Eq.~(4) and by balancing the most dominant terms,
at the  leading order one obtains \bea
\alpha=\beta=\gamma=\delta=\epsilon=\omega=-1, \eea with a set of
relations
\begin{subequations}
\bea &&p_0^2 = a_0m_0, \quad\quad\quad q_0^2 = b_0n_0,\\
&&\phi_x^2 = \frac{(c_0+c_2)}{2}\left(\sqrt{a_0b_0}+
\sqrt{m_0n_0}\right)^2.  \eea
\end{subequations}
Note that there are six functions $a_0$, $b_0$, $m_0$, $n_0$,
$p_0$ and $q_0$ (besides the arbitrary manifold $\phi(x,t)$) and
the above three conditions mean that three of them are arbitrary
at this stage of the analysis.

\subsection{Resonances}
The second step in the singularity structure analysis is to
determine the resonances (powers) at which arbitrary functions can
enter into the Laurent series (\ref{lser}). To obtain the
resonance values, we substitute the following expressions into
Eqs.~(4) \bea a &=& a_0 \phi^{-1} +\dots+ a_j \phi^{j-1}, \quad
~~~
b = b_0 \phi^{-1} +\dots+ b_j \phi^{j-1}, \nonumber\\
m &=& m_0 \phi^{-1} +\dots+ m_j \phi^{j-1}, \quad
n = n_0 \phi^{-1} +\dots+ n_j \phi^{j-1}, \nonumber\\
p &=& p_0 \phi^{-1} +\dots+ p_j \phi^{j-1}, \quad ~~~ q = q_0
\phi^{-1} +\dots+ q_j \phi^{j-1}.  \quad \eea and determine the
possible values of $j$. By collecting the coefficients of
$\phi^{j-3}$, one can obtain a system of six algebraic equations
which can be casted as
\begin{subequations}
\bea \textbf{D~X}^T &=& \textbf{0}, \eea where the superscript
`$T$' denotes
 the transpose of the matrix and the matrices \textbf{X} and \textbf{D} are given by
\bea \textbf{X} = \left(\begin{matrix} a_j & b_j & m_j & n_j & p_j
& q_j \end{matrix}\right), \eea \bea \textbf{D}= \left(
\begin{array}{cccccc}
Q_1  &  r_1~a_0^2  & r_2n_0a_0  &  r_1m_0a_0  &  2r_1a_0q_0 &  2r_1p_0a_0\\
&~&~&~& +4c_2n_0p_0 & \\
r_1~b_0^2   &   Q_1  & r_1b_0n_0  &  r_2m_0b_0  &  2r_1q_0b_0  &  2r_1p_0b_0 \\
&&&&&+4c_2m_0q_0\\
r_2b_0m_0 &  r_1a_0m_0 & Q_2  &  r_1m_0^2  &  2r_1q_0m_0 & 2r_1p_0m_0\\
&&&&+4c_2b_0p_0& \\
r_1b_0n_0 & r_2a_0n_0 & r_1n_0^2  &  Q_2     &  2r_1q_0n_0 & 2r_1p_0n_0\\
&&&&&+4c_2a_0q_0 \\
r_1b_0p_0&  r_1a_0p_0 & r_1n_0p_0 & r_1m_0p_0 & Q_3 & 2r_1a_0m_0 \\
+2c_2q_0m_0  & & +2c_2a_0q_0 & & & \\
r_1b_0q_0 & r_1a_0q_0 & r_1n_0q_0 & r_1m_0q_0 & 2r_1b_0n_0 &  Q_3\\
& +2c_2p_0n_0 & & +2c_2p_0b_0 & & \label{matrD}
\end{array}
\right),~ \eea
\end{subequations}
in which \bea
r_1 &=& c_0+c_2,\quad r_2=c_0-c_2,\nonumber\\
Q_1 &=& -(j-1)(j-2)\phi_x^2 + 2r_1~(a_0b_0+p_0q_0) + r_2~m_0n_0,\nonumber\\
Q_2 &=& -(j-1)(j-2)\phi_x^2 + 2r_1~(m_0n_0+p_0q_0) + r_2~a_0b_0,\nonumber\\
Q_3 &=& -(j-1)(j-2)\phi_x^2 + r_1~(a_0b_0+m_0n_0) +
4c_0p_0q_0\nonumber \eea and \textbf{0} is a $(6\times1)$ null
matrix. \noindent By requiring the determinant of the matrix
\textbf{D} to be zero the following resonance equation is
obtained. \bea j^3(j+1) (j-3)^3 (j-4)
\left(4c_2-3j(c_0+c_2)+j^2(c_0+c_2)\right)^2=0. \eea From Eq.~(11)
the values of $j$ are obtained as
\begin{subequations}
\bea j &=& -1 , 0 , 0 , 0 , 3 , 3 , 3 , 4 , N_1 , N_1, N_2, N_2,
\eea where \bea
N_1 &=& \frac{1}{2}\left(\frac{3(c_0+c_2)+\sqrt{9c_0^2+2c_0c_2-7c_2^2}}{c_0+c_2}\right), \\
N_2 &=&
\frac{1}{2}\left(\frac{3(c_0+c_2)-\sqrt{9c_0^2+2c_0c_2-7c_2^2}}{c_0+c_2}\right).
\eea
\end{subequations}
All the resonances should be integers for the system (2) to
satisfy the Painlev\'e property so that movable algebraic
branching type critical singular manifolds are avoided. Hence by
requiring $N_1$ and $N_2$ to be integers we find the following two
cases: \bea
\mbox{Case(i)}: \quad  c_0 &=& -\left(1+\frac{4}{m(m-3)}\right)c_2,\quad m = 1,2,4,5,6,\dots (m\neq3)\nonumber\\
\mbox{Case(ii)}: \quad    c_2 &=& 0,  \quad \quad  m = 0,3.
\nonumber \eea Then the integer resonances for both the cases can
be written as \bea j = -1 , 0 , 0 , 0 , 3 , 3 , 3 , 4 , m , m ,
3-m , 3-m, \quad  m = 0,1,2,\dots \eea Note that in the above,
$j=-1$ corresponds to the arbitrariness of the non-characteristic
manifold $\phi(x,t)$.
\newpage
\underline{\textbf{Case (i):}} \\
\indent In this case for the choices $m=1$ and $m=2$, we get
$c_0=c_2=c$ (say), where $c$ is a real constant, all the
resonances are positive integers (except for $j=-1$) and are given
below. \bea j&=&-1, 0, 0, 0, 1, 1, 2, 2, 3, 3, 3, 4. \eea

However for $m\geq4$, the presence of more negative resonances
indicates that there may exist only particular solutions with
lesser number of arbitrary functions in the Laurent expansion. For
example, the choice $m=4$ corresponding to $c_0=-2c_2$, has the
resonances $j=-1,-1,-1,0,0,0,3,3,3,4,4,4$. This may be an
indication that the Laurent expansion (5) does not correspond to a
general solution with required number of arbitrary functions, but
represents only a particular solution. We feel that these choices
with $m\geq4$ are the candidates for further deeper analysis
mathematically on Laurent expansions in the negative powers. We do
not pursue this problem further here. However, one can perform for
example
 a study on the modulation instability of system (2) for these choices of $m$($\geq 4$) and look for solitary wave type solutions which could be of
 specific physical interest. So, hereafter we will consider only the case having resonances (14) with $c_0=c_2=c$.\\
\underline{\textbf{Case (ii):}} \\
\indent For the values $m = 0$ and $m = 3$, we require  $c_2 = 0$
and the system (2) reduces to a set of three coupled nonlinear
Schr{\"o}dinger equations with resonances
$j=-1,0,0,0,0,0,3,3,3,3,3,4,$ whose integrability and Painlev\'e
analysis have already been studied in detail in Ref.~12. So we
will not consider this case any further.

\subsection{Analysis for arbitrary functions}
The next step in the Painlev\'e singularity structure analysis is
to show that there exist sufficient number of arbitrary functions
in the Laurent expansion (5) which can arise at the resonance
values given by (14) without the introduction of movable critical
singular manifolds. To prove this, we expand the dependent
variables in Eqs.~(4) (upto the highest resonance value in (14))
as below:
\begin{subequations}
\bea
a &=& \frac{a_0}{\phi}+a_1  + a_2 \phi + a_3 \phi^2 + a_4 \phi^3,\\
b &=& \frac{b_0}{\phi}+b_1  + b_2 \phi + b_3 \phi^2 + b_4 \phi^3,\\
m &=& \frac{m_0}{\phi}+m_1  + m_2 \phi + m_3 \phi^2 + m_4 \phi^3,\\
n &=& \frac{n_0}{\phi}+n_1  + n_2 \phi + n_3 \phi^2 + n_4 \phi^3,\\
p &=& \frac{p_0}{\phi}+p_1  + p_2 \phi + p_3 \phi^2 + p_4 \phi^3,\\
q &=& \frac{q_0}{\phi}+q_1  + q_2 \phi + q_3 \phi^2 + q_4 \phi^3,
\eea
\end{subequations}
where $a_j$, $b_j$, $m_j$, $n_j$, $p_j$, $q_j$, $j=0,1,\dots,4$,
are functions of $(x,t)$ to be determined. Then by collecting
various powers of $\phi$, we explicitly show that there exist
sufficient number of arbitrary functions at each index of the
resonance values given in (14). As noted above, the resonance at
$j=-1$ corresponds to the arbitrariness of the non-characteristic
manifold $\phi$.

\subsubsection{Coefficients of $\phi^{-3}$ :}
The set of algebraic equations resulting at this order is
\begin{subequations}
\bea
 - 2\phi_x^2 + 2c (a_0b_0+2p_0q_0) +  2c n_0p_0^2/a_0&=& 0, \nonumber\\
 - 2\phi_x^2 + 2c (a_0b_0+2p_0q_0) +  2c m_0q_0^2/b_0&=&0, \nonumber\\
 - 2\phi_x^2 + 2c (m_0n_0+2p_0q_0) +  2c b_0p_0^2/m_0 &=& 0,\nonumber\\
 - 2\phi_x^2 + 2c (m_0n_0+2p_0q_0) +  2c a_0q_0^2/n_0 &=&0, \nonumber\\
 - 2\phi_x^2 + 2c p_0q_0 + 2c (a_0b_0+m_0n_0) + 2c a_0q_0m_0/p_0&=& 0, \nonumber\\
 - 2\phi_x^2 + 2c p_0q_0 + 2c (a_0b_0+m_0n_0) + 2c b_0p_0n_0/q_0 &=& 0.
\label{loeq} \eea Solving Eqs.~(\ref{loeq}) again results in the
already deduced relations (8), \bea
&&p_0^2 = a_0m_0, \quad\quad q_0^2 = b_0n_0, \\
&&\phi_x^2 = c\left(\sqrt{a_0b_0}+ \sqrt{m_0n_0}\right)^2. \eea
\end{subequations}

This clearly shows that three out of the six functions ($a_0$,
$b_0$, $m_0$, $n_0$, $p_0$ and $q_0$) are arbitrary at the triple
resonance $j=0,0,0$.

\subsubsection{Coefficients of $\phi^{-2}$ :}
At the power $\phi^{-2}$, we obtain the following set of algebraic
equations expressed in the matrix form,
\begin{subequations}
\bea {\bf D_1~X_1}^T = (-i \phi_t)~{\bf Y_1}^T, \eea where \bea
{\bf D_1}&=& \left(
\begin{array}{cccccc}
4c l_1      &  2c a_0b_0  & 0         &  2c m_0n_0  &  4c l_2        &  4c p_0q_0\\
2c a_0b_0   &   4c l_1   & 2c m_0n_0  &  0          &  4c p_0q_0     &  4c l_2   \\
0           &  2c a_0b_0 &4c l_2      &  2c m_0n_0  &  4c l_1        &  4c p_0q_0  \\
2c a_0b_0   &  0         & 2c m_0n_0  &  4c l_2     &  4c p_0q_0     &  4c l_1     \\
2c l_2      &  2c a_0b_0 & 2c l_2     &  2c m_0n_0  &  2c (l_1+l_2)  &  4c p_0q_0 \\
2c a_0b_0   &  2c l_1    & 2c m_0n_0  &  2c l_2     &  4c p_0q_0
&  2c (l_1+l_2) \label{matD}
\end{array} \right),~~~~~~~\\
{\bf X_1}&=&\left(\begin{matrix} \frac{a_1}{a_0} & \frac{b_1}{b_0} & \frac{m_1}{m_0} & \frac{n_1}{n_0} & \frac{p_1}{p_0} & \frac{q_1}{q_0} \end{matrix}\right),\\
{\bf Y_1}&=&\left(\begin{matrix} ~1 & -1 & ~1 &-1 & ~1 &
-1\end{matrix}\right). \eea In the above matrix ${\bf D_1}$ we
have introduced the quantities $l_1$ and $l_2$, which are defined
as \bea
 &&l_1=a_0b_0+p_0q_0 \quad \mbox{and} \quad l_2=m_0n_0+p_0q_0.
\eea
\end{subequations}

In order to make the calculations simpler here and in the
subsequent analysis, we use the Kruskal
ansatz$^{\footnotesize{7}}$ by assuming the singular manifold
function $\phi(x,t)$ in the form $\phi(x,t) = x + \rho(t)$, with
$\rho$ an arbitrary analytic function and the $a_j$, $b_j$, $m_j$,
$n_j$, $p_j$, $q_j$ are functions of $t$ only.
One can solve the above six algebraic equations (17) given in the
matrix form and obtain
\begin{subequations}
\bea
\frac{a_1}{a_0} &=& -\frac{p_0q_0}{a_0b_0}\left(\frac{i\rho_t}{2c~l_2} +\frac{p_1}{p_0}\right),\\
\frac{b_1}{b_0} &=& ~~\frac{p_0q_0}{a_0b_0}\left(\frac{i\rho_t}{2c~l_2} -\frac{q_1}{q_0}\right),\\
\frac{m_1}{m_0} &=& -\frac{1}{l_2}\left(\frac{i\rho_t}{2c}+l_1\frac{p_1}{p_0}\right),\\
\frac{n_1}{n_0} &=&
~~~\frac{1}{l_2}\left(\frac{i\rho_t}{2c}-l_1\frac{q_1}{q_0}\right).
\eea
\end{subequations}
From equations (18), we observe that the two functions ($p_1$ and
$q_1$) out of the six functions $a_1$, $b_1$, $m_1$, $n_1$, $p_1$
and $q_1$ are arbitrary. Naturally, these are associated with the
double resonance at $j=1, 1$.

\subsubsection{Coefficients of $\phi^{-1}$ :}
At this order we obtain
\begin{subequations}
\bea {\bf D_1~X_2}^T = {\bf Y_2}^T, \eea where the matrix ${\bf
D_1}$ being defined in Eq.~(\ref{matD}) and \bea {\bf
X_2}&=&\left(
\begin{matrix} \frac{a_2}{a_0} & \frac{b_2}{b_0} & \frac{m_2}{m_0} & \frac{n_2}{n_0} & \frac{p_2}{p_0} & \frac{q_2}{q_0} \end{matrix}\right),\\
{\bf Y_2}&=& \left( \begin{matrix} y_2^{(1)} & y_2^{(2)} &
y_2^{(3)} & y_2^{(4)} & y_2^{(5)} & y_2^{(6)} \end{matrix}
\right). \eea
\end{subequations}
Here the elements of ${\bf Y_2}$ are given by
\begin{subequations}
\bea
y_2^{(1)} &=& \frac{ia_{0t}}{a_0}-\frac{2c}{a_0}\left({p_1^2n_0+a_1^2b_0}\right), \nonumber\\
&&-\frac{4 c}{a_0} \left({a_1} {p_1} {q_{0}}+ {a_1} {b_1}
{a_{0}}+{n_1} {p_1}{p_{0}}+{a_1} {q_1} {p_{0}}+{p_1}
{q_1}a_0\right),\\
y_2^{(2)} &=& -\frac{ib_{0t}}{b_0}-\frac{2c}{b_0}\left({q_1^2m_0+b_1^2a_0}\right) \nonumber\\
&&- \frac{4 c}{b_0} \left({b_1} {q_1} {p_{0}}+ {a_1} {b_1} {b_{0}}
+ {m_1} {q_1}{q_{0}} +{b_1} {p_1} {q_{0}}+ p_1{q_1}
{b_{0}}\right),\\
{y_2^{(3)}} &=& \frac{im_{0t}}{m_0}-\frac{2c}{m_0}\left({m_1^2} {n_{0}}+{p_1^2}{b_{0}}\right) \nonumber\\
&&- \frac{4 c}{m_0} \left( {m_1} {q_1} {p_{0}}+ {m_1} {n_1} {m_{0}}+{m_1} {p_1}{q_{0}} +{b_1} {p_1} {p_{0}}+{p_1} {q_1} {m_{0}} \right), \\
{y_2^{(4)}} &=& -\frac{in_{0t}}{n_0}-\frac{2c}{n_0}\left({n_1^2} {m_{0}} + {q_1^2}{a_{0}}\right) \nonumber\\
&&- \frac{4c}{n_0} \left({a_1}{q_1}{q_0}+ {m_1}{n_1}{n_0}+ {n_1}{q_1}{p_0} + {p_1}{q_1}{n_0}+ {n_1}{p_1}{q_0}\right), \\
{y_2^{(5)}} &=& \frac{ip_{0t}}{p_0}  -\frac{2c}{p_0}\left(2{p_1}
{q_1} {p_{0}}
+{a_1} {q_1} {m_{0}} +{m_1} {n_1} {p_{0}}+ {a_1} {b_1} {p_{0}}+{p_1^2} {q_{0}}\right) \nonumber\\
&&- \frac{2c}{p_0}\left({a_1} {p_1} {b_{0}} +{b_1} {p_1} {a_{0}}
+{n_1} {p_1} {m_{0}} +{a_1} {m_1} {q_{0}} +{m_1} {q_1} {a_{0}}
+{m_1} {p_1} {n_{0}}\right),~~~~~~~\\
{y_2^{(6)}} &=& -\frac{iq_{0t}}{q_0} -\frac{2c}{q_0}\left(2{p_1}
{q_1} {q_{0}}
+  {n_1} {q_1} {m_{0}} + {m_1} {n_1} {q_{0}}+  {a_1} {b_1} {q_{0}} + {q_1^2} {p_{0}}\right) \nonumber\\
&&- \frac{2c}{q_0}\left({b_1} {p_1} {n_{0}}+  {b_1} {n_1} {p_{0}}
+ {a_1} {q_1} {b_{0}}+ {m_1} {q_1} {n_{0}} + {b_1} {q_1} {a_{0}} +
{n_1} {p_1} {b_{0}}\right).~~~~ \eea
\end{subequations}
Proceeding further as in the case of $j=1$ and by incorporating
the results of $j=0$ and $j=1$, we express the four functions
$a_2$, $b_2$, $m_2$ and $n_2$ in terms of the remaining two
unknown functions $p_2$ and $q_2$:
\begin{subequations}
\bea \frac{a_2}{a_0} &=&
\frac{1}{l_1}\left(\frac{y_2^{(1)}-y_2^{(3)}}{4c}+l_2~\frac{m_2}{m_0}
-(l_2-l_1)\frac{p_2}{p_0}\right),\\
\frac{b_2}{b_0} &=&
\frac{1}{l_1}\left(\frac{y_2^{(2)}-y_2^{(4)}}{4c}+l_2~\frac{n_2}{n_0}
-(l_2-l_1)\frac{q_2}{q_0}\right), \eea where \bea \frac{m_2}{m_0}
=
\frac{n_2}{n_0}-\frac{l_1}{l_2}\left(\frac{p_2}{p_0}-\frac{q_2}{q_0}\right)
+\frac{y_2^{(3)}-y_2^{(4)}}{2c~l_2}+\frac{a_0b_0(y_2^{(1)}-y_2^{(2)}-y_2^{(3)}+y_2^{(4)})}{4~cl_1l_2},~\\
\frac{n_2}{n_0} = -\frac{l_1}{l_2}\left(\frac{q_2}{q_0}\right)
+\frac{a_0b_0(3y_2^{(4)}+y_2^{(2)}-2y_2^{(1)})+2p_0q_0(2y_2^{(4)}-y_2^{(3)})}{12~cl_1l_2},~~~~~~~~~
\eea
\end{subequations}
with $l_1$ and $l_2$ being defined in Eq.~(17e) and $y_2^{(j)}$'s,
$j=1,2,...,6$, are given in Eqs.~(20). From the above equations
(21), we can easily see that two ($p_2$ and $q_2$) out of the six
functions $a_2$, $b_2$, $m_2$, $n_2$, $p_2$ and $q_2$ are
arbitrary, as required by the existence of arbitrary functions at
the double resonance $j=2,2$.

\subsubsection{Zeroth order in $\phi$ :}
Collecting now the coefficients at the zeroth order, that is
$\phi^0$, we obtain
\begin{subequations}
\bea {\bf D_3~X_3}^T = {\bf Y_3}^T, \eea where the matrix ${\bf
D_3}$ = ${\bf D_1}- 2{\bf I}$, ~~${\bf I}$ is a $(6\times6)$
identity matrix and \bea {\bf X_3} &=& \left(
\begin{matrix} \frac{a_3}{a_0} & \frac{b_3}{b_0} & \frac{m_3}{m_0} & \frac{n_3}{n_0} & \frac{p_3}{p_0} & \frac{q_3}{q_0} \end{matrix}\right),\\
{\bf Y_3} &=& \left( \begin{matrix} y_3^{(1)} & y_3^{(2)} &
y_3^{(3)} & y_3^{(4)} & y_3^{(5)} & y_3^{(6)} \end{matrix}
\right). \eea Here \bea
y_3^{(1)} &=& i \left(\frac{a_{1t} + a_2\rho_t}{a_0}\right) - \frac{4 c}{a_0} \left({a_2} {p_1} {q_{0}} +{a_1} {a_2}{b_{0}} +{a_2} {b_1} {a_{0}}+{p_2} {q_1} {a_{0}}\right. \nonumber\\
&&\left.+{n_2} {p_1} {p_{0}} +{n_1} {p_2} {p_{0}} +{a_1} {q_2} {p_{0}} + {a_1} {p_2} {q_{0}} + {a_1} {b_2} {a_{0}}+{a_1} {p_1} {q_1} \right. \nonumber\\
&&\left.+ {a_2}{q_1}{p_{0}} + {p_1}{q_2}{a_{0}} +
{p_1}{p_2}{n_{0}}\right)
- \frac{2 c}{a_0} \left({a_1^2} {b_1} + {n_1} {p_1^2}\right),\\
y_3^{(2)} &=& -i \left(\frac{b_{1t} + b_2\rho_t}{b_0}\right) - \frac{4 c}{b_0} \left({b_1} {p_2} {q_{0}} +{b_2} {p_1}{q_{0}}+ {b_1} {p_1} {q_1}+ {p_2}{q_1}{b_{0}} \right. \nonumber\\
&&\left. + {a_2} {b_1} {b_{0}}+{p_1} {q_2} {b_{0}}
+{a_1}{b_2}{b_{0}} + {m_2}{q_1}{q_{0}} + {m_1}{q_2}{q_{0}}+ {q_1}{q_2} {m_{0}} \right. \nonumber\\
&&\left.  + {b_1}{b_2}{a_{0}} + {b_1}{q_2}{p_{0}} +
{b_2}{q_1}{p_{0}}\right)
-\frac{2 c}{b_0} \left({a_1} {b_1^2}+ {m_1} {q_1^2}\right), \\
y_3^{(3)} &=& i \left(\frac{m_{1t} + m_2\rho_t}{m_0}\right)- \frac{4 c}{m_0}  \left({p_1} {p_2} {b_{0}} + {m_1} {m_2}{n_{0}}+ {m_1} {n_2} {m_{0}}+ {m_2} {n_1} {m_{0}}  \right. \nonumber\\
&&\left. + {m_1} {p_1} {q_1} +{p_2} {q_1} {m_{0}}
+ {b_1} {p_2} {p_{0}}+ {b_2} {p_1} {p_{0}} + {p_1} {q_2} {m_{0}}+ {m_1}{p_2}{q_{0}}  \right. \nonumber\\
&&\left. + {m_2}{p_1}{q_{0}} + {m_1}{q_2}{p_{0}}
+ {m_2}{q_1}{p_{0}}\right)-\frac{2 c}{m_0} \left({b_1}{p_1^2}+{m_1^2}{n_1}\right), \\
y_3^{(4)} &=& -i \left(\frac{n_{1t} + n_2\rho_t}{n_0}\right) - \frac{4 c}{n_0} \left( n_1p_2q_0+ {n_2} {p_1} {q_{0}}+ {n_1}{q_2}{p_{0}}+ {a_1}{q_2}{q_{0}} \right. \nonumber\\
&&\left. + {m_2}{n_1}{n_{0}}+ {n_2} {q_1} {p_{0}} +{q_1} {q_2}
{a_{0}}
+ {q_1} {a_2} {q_{0}}+ {m_1} {n_2} {n_{0}}+ {p_1} {q_2} {n_{0}}\right. \nonumber\\
&&\left. + {p_2} {q_1} {n_{0}}+ {n_1} {n_2} {m_{0}} + {n_1} {p_1}
{q_1}\right) - \frac{2 c}{n_0} \left( a_1{q_1^2}+ {n_1^2} m_1
\right),\\
y_3^{(5)} &=& i \left(\frac{p_{1t} + p_2\rho_t}{p_0}\right) -
\frac{4c}{p_0} \left(p_1q_2p_0+p_2q_1p_0\right)
-\frac{2c}{p_0} \left(p_1p_2q_0+m_2p_1n_0+m_1n_1p_1 \right. \nonumber\\
&&\left. +a_1b_1p_1+a_2b_1p_0+m_1n_2p_0+m_2n_1p_0+a_1b_2p_0+b_1p_2a_0+m_1p_2n_0 \right. \nonumber\\
&&\left. +n_2p_1m_0+n_1p_2m_0+b_2p_1a_0+a_1p_2b_0+a_2p_1b_0+m_1q_2a_0 \right. \nonumber\\
&&\left.
+m_2q_1a_0+a_1m_2q_0+a_1m_1q_1+a_2m_1q_0+a_2q_1m_0+a_1q_2m_0\right),
\eea
\bea y_3^{(6)} &=& -i \left(\frac{q_{1t} + q_2\rho_t}{q_0}\right)
- \frac{4c}{q_0} \left(p_1q_2q_0+p_2q_1q_0\right)
-\frac{2c}{p_0} \left(q_1q_2p_0+m_2q_1n_0+m_1n_1q_1 \right. \nonumber\\
&&\left. +a_1b_1q_1+b_1q_2a_0+m_1q_2n_0+n_2q_1m_0+n_1q_2m_0+b_2q_1a_0+m_1n_2q_0  \right. \nonumber\\
&&\left. +a_2b_1q_0+m_2n_1q_0+a_1b_2q_0+b_1n_2p_0+a_1q_2b_0+a_2q_1b_0  \right. \nonumber\\
&&\left.
+b_2n_1p_0+n_2p_1b_0+b_1n_1p_1+n_1p_2b_0+b_2p_1n_0+b_1p_2n_0\right).
\eea
\end{subequations}
After a straightforward but lengthy algebra one can solve the
above Eqs.~(22) and deduce the following three expressions:
\begin{subequations}
\bea \frac{a_3}{a_0} &=& \frac{\left( (4cl_2-2)\frac{m_3}{m_0}
-(4c(l_2-l_1)-2)\frac{p_3}{p_0}+y_3^{(1)}-y_3^{(3)}\right)}{(4cl_1-2)}, \\
\frac{b_3}{b_0} &=& \frac{\left( (4cl_2-2)\frac{n_3}{n_0}
-(4c(l_2-l_1)-2)\frac{q_3}{q_0}+y_3^{(2)}-y_3^{(4)}\right)}{(4cl_1-2)},
\eea where \bea \frac{m_3}{m_0} =
-\frac{n_3}{n_0}+\frac{\left({4cl_1}\left(\frac{p_3}{p_0}+\frac{q_3}{q_0}\right)-y_3^{(3)}\right)}{2c(l_1-l_2)}
+\frac{a_0b_0(y_3^{(2)}-y_3^{(4)})}{2c(l_1-l_2)^2}. \eea
\end{subequations}
The above expressions (23) indicate the arbitrariness of three
functions ($n_3$, $p_3$ and $q_3$) out of the six functions $a_3$,
$b_3$, $m_3$, $n_3$, $p_3$ and $q_3$. Thus system (2) with
$c_0=c_2=c$, satisfies the requirement of the presence of three
arbitrary functions corresponding to the triple resonance at
$j=3,3,3$.

\subsubsection{Coefficients of $\phi^1$ :}
In a similar manner, after a lengthy algebra carried out using
\emph{Maple} we have verified that the resulting six algebraic
equations at the coefficient of $\phi$, which we do not present
here for want of space, reduce to five equations with six unknown
functions. Thus we observe that there exists one arbitrary
function corresponding to the resonance $j=4$, as required.

Our preceding analysis shows that for the choice $c_0=c_2=c$,
there exist sufficient number of arbitrary functions at the
resonance values given by Eq.~(14). One can proceed further to
obtain the higher order coefficient functions for all $j>4$ in
terms of the previous coefficients without the introduction of any
movable critical singular manifold into the Laurent expansion for
the case $c_0=c_2=c$. So we conclude that the system of three
component Gross-Pitaevskii (GP) type equation (2) passes the
Painlev\'e test only for the two cases (i) $c_2=0$ and (ii)
$c_0=c_2=c$ and is expected to be integrable. Of course, this fact
has already been shown for the case $c_2=0$ through the Painlev\'e
analysis$^{\footnotesize{12}}$ and the second case $c_0=c_2=c$ can
be reduced to the $(2\times2)$ matrix NLS
equation$^{\footnotesize{28}}$ which is integrable through the IST
method$^{\footnotesize{33,34}}$.

\section{Conclusion}
In this paper, we have studied the integrability property of the
three component Gross-Pitaevskii (GP) type equations arising as
the evolution equations for spinor condensates by applying the
Painlev\'e singularity structure analysis. We have identified that
only for the following two choices of the effective one
dimensional coupling constants, (i) $c_2=0$ and (ii) $c_0=c_2=c$,
the system (2) passes the Painlev\'e test and possesses Laurent
expansion with full complement of arbitrary functions without the
introduction of movable critical singular manifolds. The
integrability of the first choice ($c_2=0$) has been discussed
already$^{\footnotesize{12}}$. For the second choice ($c_0=c_2=c$)
by applying the IST method the multi-bright solitons under
vanishing$^{\footnotesize{28,29}}$ as well as non-vanishing
boundary conditions$^{\footnotesize{31}}$ and multi-dark
solitons$^{\footnotesize{30}}$ have been obtained by reducing the
system (2) to the known IST integrable $(2\times2)$ matrix
nonlinear Schr\"odinger equation$^{\footnotesize{33,34}}$. Our
present analysis also shows that the system (2) passes the
Painlev\'e test for integrability when $c_0$ and
$c_2$ are equal and non-zero, in addition to the choice $c_2=0$. 
Apart from finding the choices for which the system (2) can be
integrable, one can also obtain information regarding the Hirota's
bilinearization for such cases from the above analysis. This has
already been exploited for the case $c_2=0$ to obtain
multi-soliton solutions$^{\footnotesize{35,36}}$. Work is now in
progress to make a similar analysis for the case $c_0=c_2=c$. Also
it is of future interest to identify multicomponent integrable
systems with higher degree of hyperfine spin ($F>1$), see for
example the coupled evolution equations given
in$^{\footnotesize{37}}$, for which also the above type of
Painlev{\'e} singularity structure analysis can be carried out.

\section*{\large{Acknowledgements}}
T. K. acknowledges the support of Department of Science and
Technology, Government of India under the DST Fast Track Project
for young scientists.  T. K. and K. S. thank the Principal and
Management of Bishop Heber College, Tiruchirapalli, for constant
support and encouragement.  The work of M. L. is supported by a
DST-IRPHA project and DST Ramanna Fellowship.

\begin{tabular}{p{.15cm}p{14cm}}
\footnotesize$^1$ &
M.~Lakshmanan and S.~Rajasekar, \emph{Nonlinear Dynamics:
Integrability, Chaos and Patterns} (Springer-Verlag, New York,
2003).\\ \footnotesize$^2$ &
M.~J.~Ablowitz and P.~A.~Clarkson, \emph{Solitons, Nonlinear
Evolution Equations and Inverse Scattering} (Cambridge University
Press, Cambridge, 1992).\\
\footnotesize$^3$ &
Y.~S.~Kivshar and G.~P.~Agrawal,~{\it  Optical Solitons: From
Fibers to Photonic Crystals} (Academic Press, San Diego, 2003).\\
\footnotesize$^4$ &
T.~Dauxois and M.~Peyrad, \emph{Physics of Solitons} (Cambridge
University Press, Cambridge, 2006).\\
\footnotesize$^5$ &
A.~C.~Scott, {\it Nonlinear Science: Emergence and Dynamics of
Coherent Structures} (Oxford University Press, Oxford, 1999).\\
\footnotesize$^6$ &
M.~J.~Ablowitz, A.~Ramani, and H.~Segur, {\it J.~Math.~Phys.}
\textbf{21}, 715 (1980).\\
\footnotesize$^7$ &
M.~Jimbo, M.~D.~Kruskal, and T.~Miwa, {\it Phys.~Lett.~A}
\textbf{92}, 59 (1982).\\
\footnotesize$^8$ &
J.~Weiss, M.~Tabor, and G.~Carnevale,  {\it J.~Math.~Phys.}
\textbf{24}, 522 (1983).\\
\footnotesize$^9$ &
J.~Weiss, {\it J.~Math.~Phys.} \textbf{25}, 13 (1984).\\
\footnotesize$^{10}$ &
M.~Lakshmanan and R.~Sahadevan, {\it Phys.~Rep.}~\textbf{224},
1 (1993).\\

\footnotesize$^{11}$ &
R.~Sahadevan, K.~M.~Tamizhmani, and M.~Lakshmanan, {\it
J.~Phys.~A}
\textbf{19}, 1783 (1986).\\
\footnotesize$^{12}$ &
R.~Radhakrishnan, R.~Sahadevan, and M.~Lakshmanan, {\it Chaos,
Solitons~and~Fractals} \textbf{5}, 2315 (1995).\\
\footnotesize$^{13}$ &
R.~Radhakrishnan, M.~Lakshmanan, and M.~Daniel, {\it J.~Phys.~A}
\textbf{28}, 7299 (1995).\\
\footnotesize$^{14}$ &
Q-Han Park and H.~J.~Shin, {\it Phys.~Rev.~E} \textbf{59},
2373 (1999).\\
\footnotesize$^{15}$ &
S.~Yu.~Sakovich and T.~Tsuchida, {\it J.~Phys.~A:~Math.~Gen.}
\textbf{33}, 7217 (2000).\\
\footnotesize$^{16}$ &
D.~Schumayer and B.~Apagyi, {\it J.~Phys.~A:~Math.~Gen.}
\textbf{34}, 4969 (2001).\\

\footnotesize$^{17}$ &
Tin-Lun~Ho and V.~B.~Shenoy, {\it Phys.~Rev.~Lett.} \textbf{77},
3276 (1996).\\
\footnotesize$^{18}$ &
R.~J.~Ballagh, K.~Burnett, and T.~F.~Scott, {\it Phys.~Rev.~Lett.}
\textbf{78}, 1607 (1997).\\
\footnotesize$^{19}$ &
S.~Rajendran, P.~Muruganandam, and M.~Lakshmanan, {\it J.~Phys.~B:
At.~Mol.~Opt.~Phys.} {\bf 42}, 145307 (2009).\\
\footnotesize$^{20}$ &
J.~Stenger, S.~Inouye, D.~M.~Stamper-Kurn, H.~J.~Miesner,
A.~P.~Chikkatur, and W.~Ketterle, {\it Nature (London)}
\textbf{396},
345 (1998).\\
\end{tabular}

\newpage
\begin{tabular}{p{.15cm}p{14cm}}
\footnotesize$^{21}$ &
D.~M.~Stamper-Kurn, M.~R.~Andrews, A.~P.~Chikkatur, S.~Inouye, H.~
J.~Miesner, J.~Stenger, and W.~Ketterle, {\it Phys.~Rev.~Lett.}
\textbf{80}, 2027 (1998).\\
\footnotesize$^{22}$ &
H.~J.~Meisner, D.~M.~Stamper-Kurn, J.~Stenger, S.~Inouye,
A.~P.~Chikkatur,~and W.~Ketterle, {\it Phys.~Rev.~Lett.}
\textbf{82},
2228 (1999).\\
\footnotesize$^{23}$ &
C.~J.~Myatt, E.~A.~Burt, R.~W.~Ghrist, E.~A.~Cornell, and C.~E.~Wieman,~{\it Phys.~Rev.~Lett.} \textbf{78}, 586 (1997).\\
\footnotesize$^{24}$ &
Tin-Lun~Ho, {\it Phys.~Rev.~Lett.} \textbf{81}, 742 (1998).\\
\footnotesize$^{25}$ &
T.~Ohmi and K.~Machida, {\it J.~Phys.~Soc.~Jpn.}~\textbf{67},
1822~(1998).\\
\footnotesize$^{26}$ &
E.~V.~Goldstein and P.~Meystre, {\it Phys. Rev.~A} \textbf{59},
1509 (1999).\\
\footnotesize$^{27}$ &
C.~K.~Law, H.~Pu, and N.~P.~Bigelow, {\it Phys.~Rev.~Lett.}
\textbf{81}, 5257 (1998).\\
\footnotesize$^{28}$ &
J.~Ieda, T.~Miyakawa, and M.~Wadati, {\it Phys.~Rev.~Lett.}
\textbf{93}, 194102 (2004).\\
\footnotesize$^{29}$ &
J.~Ieda, T.~Miyakawa, and M.~Wadati, {\it J.~Phys.~Soc.~Jpn.}
\textbf{73}, 2996 (2004).\\
\footnotesize$^{30}$ &
M.~Uchiyama, J.~Ieda, and M.~Wadati, {\it J.~Phys.~Soc.~Jpn.}
\textbf{75}, 064002 (2006).\\
\footnotesize$^{31}$ &
T.~Kurosaki and M.~Wadati, {\it J.~Phys.~Soc.~Jpn.} \textbf{76},
084002 (2007).\\
\footnotesize$^{32}$ &
M.~Daniel, M.~D.~Kruskal, M.~Lakshmanan, and K.~Nakamura, {\it J.~Math.~Phys.} \textbf{33}, 771 (1992).\\
\footnotesize$^{33}$ &
T.~Tsuchida and M.~Wadati, {\it J.~Phys.~Soc.~Jpn.} {\bf 67},
1175 (1998).\\
\footnotesize$^{34}$ &
J.~Ieda, M.~Uchiyama, and M.~Wadati, {\it J.~Math.~Phys.} {\bf
48},
013507 (2007).\\
\footnotesize$^{35}$ &
T.~Kanna and M.~Lakshmanan, {\it Phys.~Rev.~Lett.} {\bf 86},  5043
(2001).\\
\footnotesize$^{36}$ &
T.~Kanna and M.~Lakshmanan, {\it Phys.~Rev.~E} {\bf 67},  046617
(2003).\\
\footnotesize$^{37}$ &
M.~Uchiyama, J.~Ieda, and M.~Wadati, {\it J.~Phys.~Soc.~Jpn.} {\bf
76}, 074005 (2007).\\

\end{tabular}
\end{document}